%% file: skeleton.tex
\documentclass[a4paper,11pt]{article}
\usepackage{pos}
\usepackage{amsmath}

\usepackage{amssymb}
\usepackage{anyfontsize}
\usepackage{subfigure}

\DeclareMathAlphabet{\mathcal}{OMS}{cmsy}{m}{n}

\graphicspath{{./}}

\input{newcommands.tex}

\title{Mass and isovector matrix elements of the nucleon at zero-momentum transfer}

\author*[a]{Konstantin~Ottnad}
\author[b,c]{Dalibor~Djukanovic}
\author[a,b,c]{Harvey~B.~Meyer}
\author[a]{Georg~von~Hippel}
\author[a,b,c]{Hartmut~Wittig}

\affiliation[a]{PRISMA\textsuperscript{+} Cluster of Excellence and Institut f\"ur Kernphysik, Johannes Gutenberg-Universität Mainz, \\
Johann-Joachim-Becher-Weg 45, 55099 Mainz, Germany}

\affiliation[b]{Helmholtz Institute Mainz, Johannes Gutenberg-Universität Mainz, 55099 Mainz, Germany}

\affiliation[c]{GSI Helmholtzzentrum für Schwerionenforschung, Planckstraße 1, 64291 Darmstadt, Germany}

\emailAdd{kottnad@uni-mainz.de}

\abstract{We present the current status of our analysis of nucleon structure observables including isovector charges and twist-2 matrix elements as well as the nucleon mass. Results are computed on a large set of CLS $N_f=2+1$ gauge ensembles with $M_\pi\approx0.130\mev\ldots 350\mev$, four values of the lattice spacing $a=0.05\fm\ldots 0.09\fm$ and covering a large range of physical volumes. Compared to the results presented at last year's conference we have added data on a very fine and large box at small light quark mass ($T\times L^3=192 \times 96^3$, $M_\pi=172\mev$, $a=0.05\fm$). Additional (intermediate) source-sink separations have been computed on the coarser ensembles, further increasing effective statistics and allowing for a more fine-grained control in the treatment of the excited-state contamination. Excited states in the nucleon matrix elements are tamed by a simultaneous, two-state fit ansatz using the summation method. \\[3em] \flushright MITP-22-107}

\FullConference{%
  The 39th International Symposium on Lattice Field Theory (Lattice2022),\\
  8-13 August, 2022 \\
  Bonn, Germany 
}

\begin{document}
\maketitle

\section{Introduction}
The experimentally precisely known isovector axial charge of the nucleon has long been considered a benchmark for nucleon structure calculations in lattice QCD \cite{FlavourLatticeAveragingGroupFLAG:2021npn}. Even though a lot of progress has been made in recent years, such calculations remain challenging due to the inherent signal-to-noise problem and the resulting excited-state contamination of ground state matrix elements. Nucleon charges or, more generally, nucleon matrix elements (NMEs) at zero-momentum transfer are at least in principle straightforward to compute and among the statistically most precise nucleon structure observables on the lattice. Therefore, they are not only of phenomenological interest but also provide a fairly clean testbed for various techniques to tame excited-state contaminations. \par

In a proceedings contribution for last year's conference in Ref.~\cite{Ottnad:2021tlx} we have first presented an effort to update our results on isovector NMEs at zero-momentum transfer in Ref.~\cite{Harris:2019bih}. Here we report on the continuation and status of this work concerning the isovector NMEs, but also include a dedicated analysis of the nucleon mass $M_N$ measured on the same set of gauge configurations providing a consistency check for the scale setting procedure that has been used in our calculations. The details of our lattice setup including the computation of the two- and three-point functions are given in section~\ref{sec:setup}, whereas the fitting procedure and physical extrapolation of the nucleon mass is discussed in section~\ref{sec:M_N}. Our excited-state analysis and strategy for physical extrapolation of the NMEs are explained in section~\ref{sec:NME}. Lastly, a summary and outlook towards a final analysis is included in section~\ref{sec:outlook}. \par

\section{Lattice Setup} \label{sec:setup}
The ensembles used in this study have been generated by the Coordinated Lattice Simulation (CLS) initiative with $N_f=2+1$ dynamical quark flavors of Wilson fermions and the tree-level improved Symanzik gauge action as described in Ref.~\cite{Bruno:2014jqa}. A twisted mass regulator is used to suppress exceptional configurations \cite{Luscher:2012av}. Most ensembles have been simulated with open boundary conditions (oBC) in time to counter topological charge freezing \cite{Luscher:2011kk} with the exception of a few ensembles at $a>0.05\fm$ with periodic boundary conditions (pBC). An overview of the ensemble included in this study can be found in Table~\ref{tab:ensembles}. Data for E300 and N101 have been newly added and had not yet been available in Ref.~\cite{Ottnad:2021tlx}. The addition of E300 yields another data point close to physical quark masses ($M_\pi\approx173\mev$) at the finest lattice spacing, further constraining the chiral and continuum extrapolation. On the other hand, N101 has the largest value of $M_\pi L$ among all ensembles and a large physical volume while otherwise sharing its input parameters with H105, thus improving finite volume extrapolations. \par

\begin{table}[!t]
 \caption{Gauge ensembles used in this work. Ensembles with open and periodic boundary conditions in time are indicated by superscripts ``$o$'' and ``$p$'', respectively. Lattice spacings are taken from Ref.~\cite{Bruno:2014jqa}. $M_\pi$ and $M_N$ have been measured on the same set of configurations and the corresponding values of $M_\pi L$ are included as well. $N_\mathrm{conf}$ is the number of gauge configurations measurements, and $N_\mathrm{meas}^\mathrm{max}$ refers to the number of measurements on the largest value of $\tsep$. The range of source-sink separations is given in physical units by $\tseplo$ and $\tsephi$, and $N_{\tsep}$ is the number of equidistant source-sink separations on each ensemble.}
 \centering
 \setlength{\tabcolsep}{0.28em}
 \begin{tabular*}{\textwidth}{crcccccccccc}
  \hline\hline
  ID$^\mathrm{BC}$ & $\frac{T}{a}\times(\frac{L}{a})^2$ & $a/\fm$ & $M_\pi L$ & $M_\pi/\mev$ & $M_N/\mev$ & $N_\mathrm{conf}$ & $N_\mathrm{meas}^\mathrm{max}$ & $\tseplo/\fm$ & $\tsephi/\fm$ & $N_{\tsep}$ \\
  \hline\hline
  C101$^o$ &  $96 \times 48^3$ & 0.08636 & 4.7 & 224(3) &  974(12) & 2000 &  64000 & 0.35 & 1.47 & 14 \\
  N101$^o$ & $128 \times 48^3$ &         & 5.9 & 280(3) & 1022(12) & 1595 &  51040 &      &      & 14 \\
  H105$^o$ &  $96 \times 32^3$ &         & 3.9 & 279(4) & 1029(15) & 1027 &  49296 &      &      & 14 \\
  H102$^o$ &  $96 \times 32^3$ &         & 5.0 & 352(4) & 1093(13) & 2037 &  32592 &      &      & 14 \\
  \hline
  D450$^p$ & $128 \times 64^3$ & 0.07634 & 5.4 & 215(2) &  974(11) &  500 &  64000 & 0.31 & 1.53 & 17 \\
  N451$^p$ & $128 \times 48^3$ &         & 5.3 & 285(3) & 1039(12) & 1011 & 129408 &      &      &  9 \\
  S400$^o$ & $128 \times 32^3$ &         & 4.3 & 348(4) & 1117(13) & 2873 &  45968 &      &      &  9 \\
  \hline
  E250$^p$ & $192 \times 96^3$ & 0.06426 & 4.0 & 128(2) &  927(11) &  400 & 102400 & 0.26 & 1.41 & 10 \\
  D200$^o$ & $128 \times 64^3$ &         & 4.2 & 202(2) &  961(12) & 1999 &  63968 &      &      & 10 \\
  N200$^o$ & $128 \times 48^3$ &         & 4.4 & 280(3) & 1058(13) & 1712 &  20544 &      &      & 10 \\
  S201$^o$ & $128 \times 32^3$ &         & 3.0 & 291(4) & 1120(14) & 2093 &  66976 &      &      & 10 \\
  N203$^o$ & $128 \times 48^3$ &         & 5.4 & 344(4) & 1107(13) & 1543 &  24688 &      &      & 10 \\
  \hline
  E300$^o$ & $192 \times 96^3$ & 0.04981 & 4.2 & 173(2) &  962(12) &  569 &  18208 & 0.20 & 1.40 & 13 \\
  J303$^o$ & $192 \times 64^3$ &         & 4.2 & 258(3) & 1034(12) & 1073 &  17168 &      &      & 13 \\
  N302$^o$ & $128 \times 48^3$ &         & 4.2 & 347(4) & 1140(13) & 2201 &  35216 &      &      & 13 \\
  \hline\hline
 \end{tabular*}
 \label{tab:ensembles}
\end{table}

Our analysis requires the computation of quark-connected two- and three-point functions
\begin{align}
 C^\mathrm{2pt}(\Gamma; t) &= \sum_{\vec{x}} \Gamma_{\alpha\beta} \langle N_\alpha(\vec{x}, t) \bar{N}_\beta(0)\rangle \label{eq:2pt} \,, \\
 C^X_{\mu_1...\mu_n}(\Gamma; \tins, \tsep) &= \sum_{\vec{x},\vec{y}} \Gamma_{\alpha\beta} \langle N_\alpha(\vec{x}, \tsep) \mathcal{O}^X_{\mu_1...\mu_n}(\vec{y}, \tins) \bar{N}_\beta(0) \rangle \label{eq:3pt} \,,
\end{align}
where $N_\alpha(\vec{x}, t)$ denotes the nucleon interpolating field, $\mathcal{O}^X_{\mu_1...\mu_n}(\vec{y}, \tins)$ the operator insertion and the actual choice of the spin projector $\Gamma$ depends on the analysis, cf. sections~\ref{sec:M_N}~and~\ref{sec:NME}. Initial and final state are always produced at rest and only vanishing momentum transfer is required for $C^X_{\mu_1...\mu_n}(\Gamma; \tins, \tsep)$. The two relevant Euclidean time-separations for $C^X_{\mu_1...\mu_n}(\Gamma; \tins, \tsep)$ are the source-sink separation $\tsep=t_f-t_i$ and the insertion time $\tins=t - t_i$, where $t_i$, $t$ and $t_f$ are the time positions of initial state, operator insertion and final state. The two- and three-point functions are computed on a common set of point sources using the truncated solver method \cite{Bali:2009hu,Blum:2012uh,Shintani:2014vja} and with the same smearing setup as described in Ref.~\cite{Harris:2019bih}. Since (part of) the data are used for other projects as well, e.g. the computation of electromagnetic and axial form factors \cite{Djukanovic:2019jtp,Djukanovic:2021cgp,Djukanovic:2022wru} also non-vanishing momenta have been computed for both $C^\mathrm{2pt}(\Gamma; t)$ and $C^X_{\mu_1...\mu_n}(\Gamma; \tins, \tsep)$.\par

The source setup depends on the choice of boundary conditions. For ensembles with pBC sources are distributed randomly on the entire lattice volume. $N_\mathrm{meas}^\mathrm{max}$ in Table~\ref{tab:ensembles} is the resulting total number of measurements at the largest value of $\tsep$. Going to smaller values of $\tsep$ the number of measurements $N_\mathrm{meas}$ is divided in half every one or two steps in $\tsep$. This scaling of measurements is chosen such that the behavior of effective statistics is closer to a constant in $\tsep$ instead of showing an exponential decay of the signal-to-noise ratio for increasing values of $\tsep$. On ensembles with oBC the sources are distributed on a single time slice in the bulk of the lattice and the scaling of measurements is only applied for $\tsep\lesssim 1fm$ on most ensembles, as the data for $\tsep\gtrsim 1\fm$ had been produced for the older study in Ref.~\cite{Harris:2019bih} without any scaling of statistics. \par

The renormalization factors for the NMEs are the same as the ones that have been used in Ref.~\cite{Harris:2019bih}. We use the gradient flow scale $t_0/a^2$ introduced in Ref.~\cite{Luscher:2010iy} to form dimensionless quantities from any observable before carrying out the physical extrapolation. For the values for $t_0/a^2$, the determination of the physical value $\sqrt{8t_0}=0.415\stat{4}\sys{2}$ that we use to set the scale, as well as further details on the scale setting procedure the reader is referred to Ref.~\cite{Bruno:2016plf}. \par

\section{Nucleon mass} \label{sec:M_N}
\begin{figure}[thb]
 \centering
 \includegraphics[totalheight=0.275\textheight]{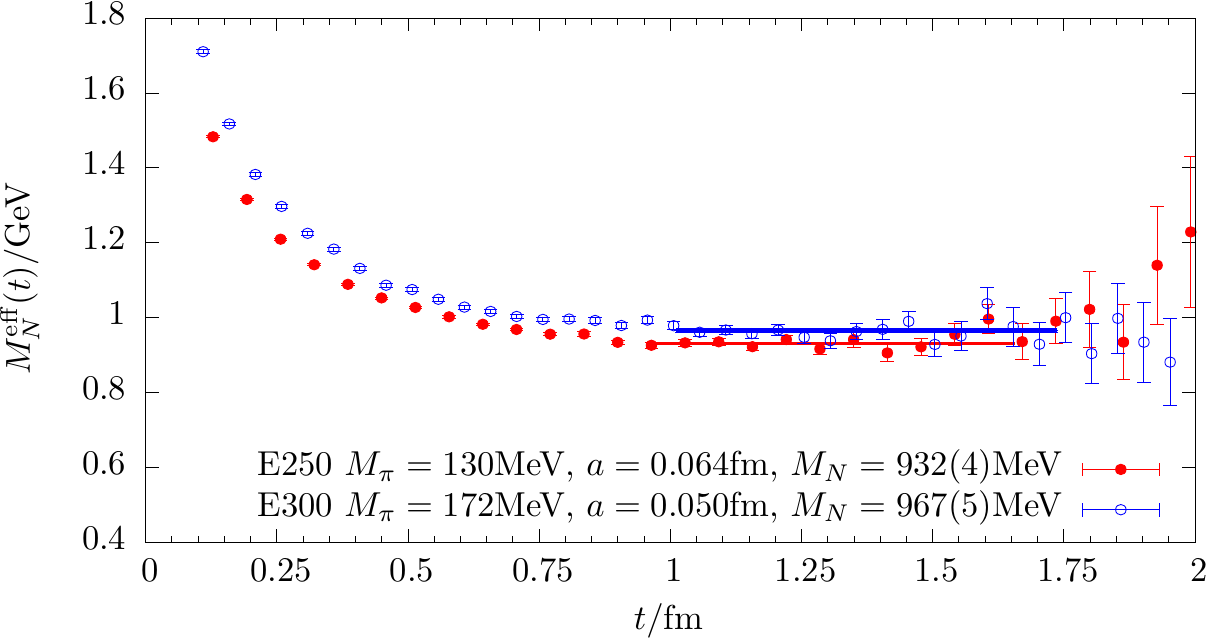}
 \caption{Effective nucleon mass on the two most chiral ensembles E250 and E300. Fitted values, errors and ranges are indicated by the bands. Errors are statistical only and do not contain the scale setting error from the conversion to physical units.}
 \label{fig:eff_mass_and_m_N_chiral_extrapolation}
\end{figure}

The nucleon mass has been computed on each ensemble from a single exponential fit to the gauge averaged two-point function in Eq.~\ref{eq:2pt}. Use of the unpolarized spin projector $\Gamma=\Gamma_0\equiv\frac{1}{2}\l(1+\g{0}\r)$ leads to a favorable signal-to-noise ratio compared to a single, polarized projector. An example of the data quality for the effective nucleon mass on our two most chiral ensembles is shown in Fig.~\ref{fig:eff_mass_and_m_N_chiral_extrapolation} together with the results from the fit to the corresponding correlation function. The relative statistical errors on the fitted values of $M_N$ are typically a few permille on most ensembles. \par

\begin{figure}[thb]
 \centering
 \includegraphics[totalheight=0.196\textheight]{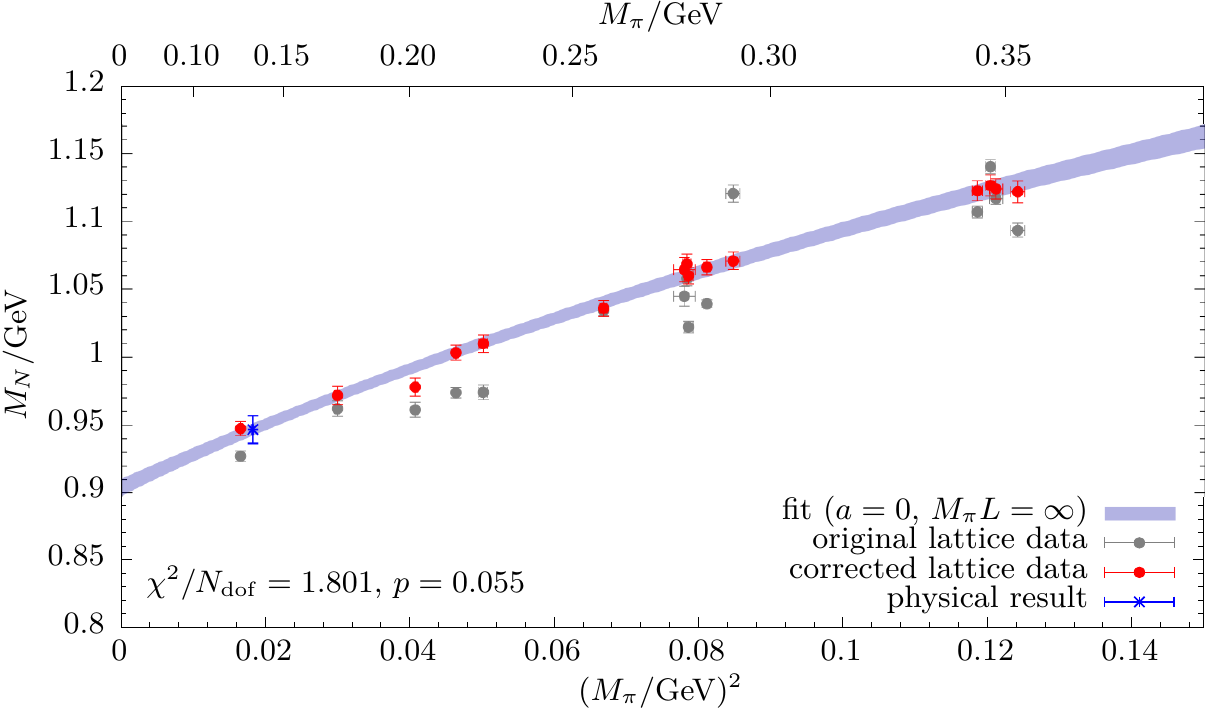}
 \includegraphics[totalheight=0.196\textheight]{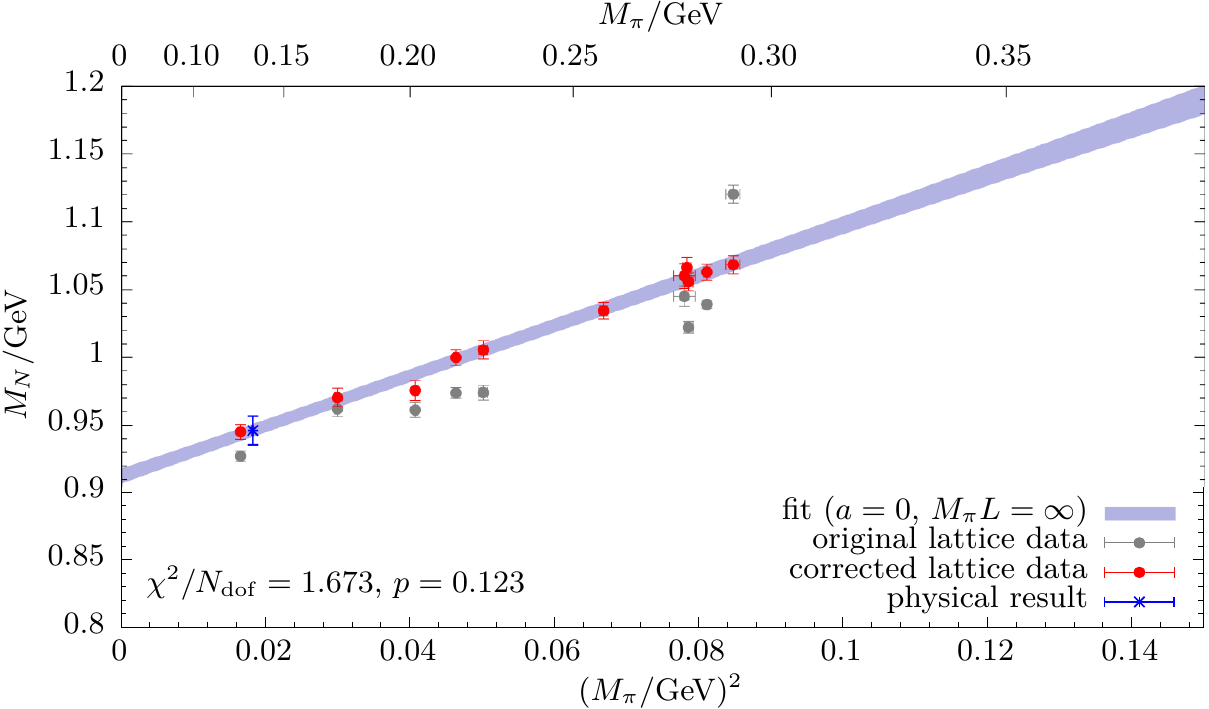} \\[3pt]
 \includegraphics[totalheight=0.1925\textheight]{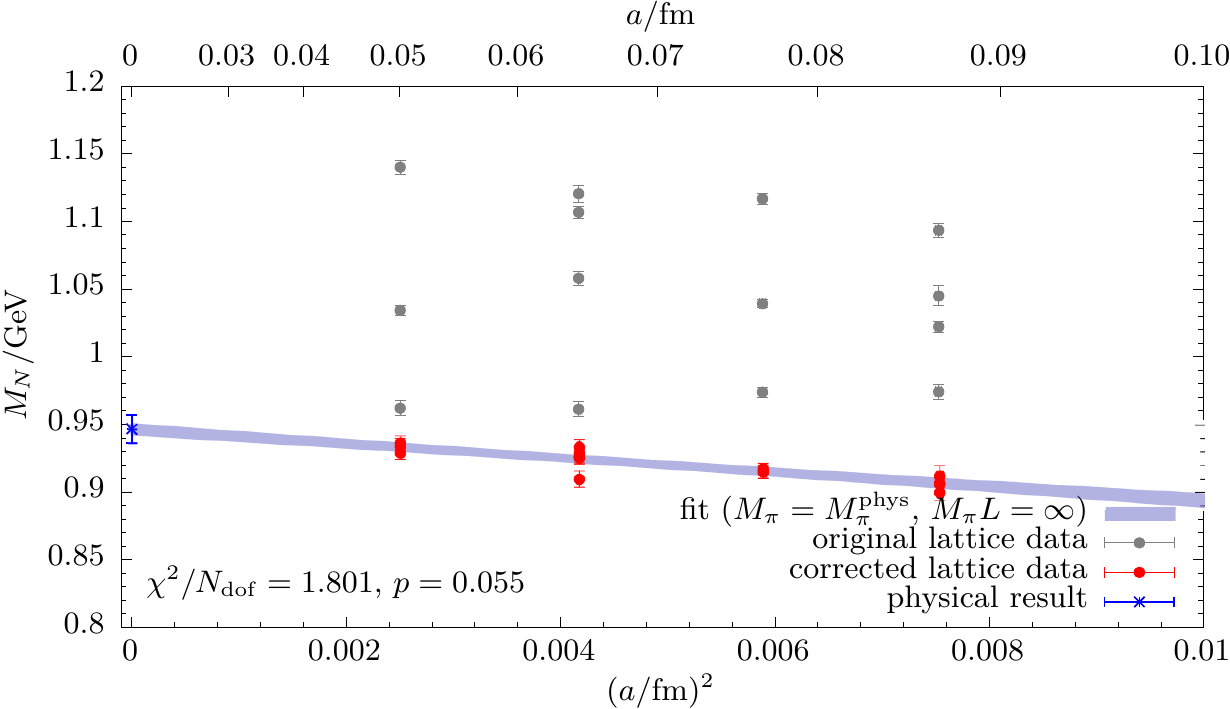}
 \includegraphics[totalheight=0.177\textheight]{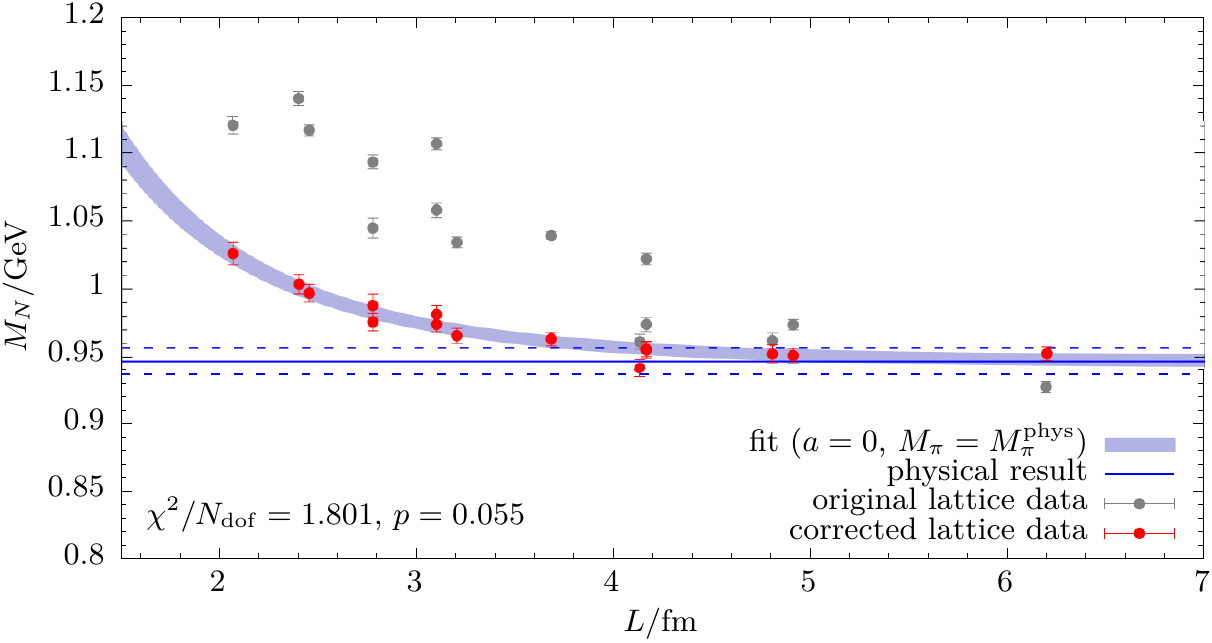}
 \caption{Physical extrapolation of the nucleon mass. The upper row shows the chiral extrapolation for the full set of ensembles (left panel) and with a cut of $M_\pi<300\mev$ (right panel). Continuum and finite volume extrapolations for the full set of ensembles are displayed in the lower two panels. The red data points are obtained from the original (gray) lattice data by correcting for the physical extrapolation in the variables not shown in any given plot (e.g. continuum and infinite volume limit in case of the upper two panels) using the parameters from the fit. Therefore the errors on the red points are strongly correlated. Errors on individual data points and the width of the filled, blue extrapolation bands are statistical only, whereas the error on the physical result includes the scale setting error from conversion to physical units.}
 \label{fig:m_N_CCF_extrapolations}
\end{figure}

The physical extrapolation is carried out using a $\chi$PT-inspired fit ansatz~\cite{Schindler:2006ha}
\begin{equation}
 M_N(M_\pi, a, L) = \chiral{M}_N + B M_\pi^2 + C M_\pi^3 + D a^2 + E \frac{M_\pi^3}{(M_\pi L)} e^{-M_\pi L} \,,
 \label{eq:CCF_M_N}
\end{equation} 
where the term $\sim D a^2$ accounts for the leading lattice artifact and the last one $\sim E \frac{M_\pi^3}{(M_\pi L)} e^{-M_\pi L}$ for finite volume effects, cf. Ref.~\cite{Beane:2004tw}. We find that including further terms of $\mathcal{O}(M_\pi^4)$ does not improve the fit as the corresponding free parameters are unconstrained by our lattice data at the current level of precision. The fits are performed using a binned bootstrap with $N_B=10000$ bootstrap samples to propagate statistical errors from the input quantities to the final result. Fitting the full set of available lattice data, we find a physical value of the nucleon mass of
\begin{equation}
 M_N^\phys = 947(10) \mev \,,
\end{equation}
compatible with the experimental value. The physical pion mass $M_\pi^\phys$ is fixed in units of $\sqrt{t_0}$ using its value in the isospin limit, i.e. $M_\pi^\phys=M_\pi^\mathrm{iso}=134.8(3)\mev$~\cite{Aoki:2016frl}, and the uncertainty due to the use of $\sqrt{t}_0$ is propagated in the extrapolation. However, the uncertainty from fixing the physical point in $M_\pi$ is much smaller than the one from the conversion of $\sqrt{t_0} M_N^\phys$ to physical units, which entirely dominates the error of the final result. This is also visible in Fig.~\ref{fig:m_N_CCF_extrapolations}, where the width of the extrapolation bands from the resulting chiral, continuum and finite volume extrapolations does not contain the error from the conversion, unlike the larger error on the physical result.

In order to assess the reliability of the physical extrapolation we have applied three different cuts to the input data, i.e. $M_\pi<300\mev$, $a<0.075\fm$ and $M_\pi L>4$, leading to
$M_N^\phys=946(11)\mev$, $M_N^\phys=945(11)\mev$ and $M_N^\phys=949(10)\mev$, respectively. Whereas these results are all in excellent agreement within errors, there is some mild tension for the resulting slopes of the chiral extrapolation with and without a cut in $M_\pi$ as depicted in the upper two panels of Fig.~\ref{fig:m_N_CCF_extrapolations}, i.e. the fit is essentially unable to resolve terms beyond $\sim M_\pi^2$ when leaving out the ensembles at $M_\pi\geq300\mev$. \par

\section{Matrix elements} \label{sec:NME}
The forward NMEs considered in this study are of the form
\begin{equation}
 \bra{N(s_f)} \mathcal{O}^X_{\mu_1...\mu_n} \ket{N(s_i)} = \bar{u}(s_f) W^X_{\mu_1...\mu_n}  u(s_i) \,,
 \label{eq:NMEs}
\end{equation}
where $u(s_i)$, $\bar{u}(s_f)$ denote Dirac spinors with initial (final) state spin $s_i$ ($s_f$). Operator insertions $\mathcal{O}^X_{\mu_1...\mu_n}$ are restricted to quark isovector combinations only. Six different operators are considered that can be divided into two groups, i.e. local, dimension-three operators
\begin{equation}
 \mathcal{O}^A_\mu=\bar{q} \gamma_\mu \gamma_5 q, \quad \mathcal{O}^S=\bar{q}q , \quad \mathcal{O}^T_{\mu\nu} = \bar{q} i \sigma_{\mu\nu} q \,, 
 \label{eq:local_operators}
\end{equation}
and dimension-four operators
\begin{equation}
 \mathcal{O}^{vD}_{\mu\nu}=\bar{q} \g{\l\{\mu\r.} \DBF{\l.\nu\r\}} q \,, \qquad \mathcal{O}^{aD}_{\mu\nu}=\bar{q} \g{\l\{\mu\r.} \g{5} \DBF{\l.\nu\r\}} q \,, \qquad \mathcal{O}^{tD}_{\mu\nu\rho}=\bar{q} \sigma_{\l[\mu\l\{\nu\r.\r]} \DBF{\l.\rho\r\}} q \,,
 \label{eq:twist2_operators}
\end{equation}
that exhibit one (symmetric) derivative $\,\DBF{\mu}=\frac{1}{2} (\DF{\mu}-\DB{\mu})$. The notation $\{...\}$, $[...]$ refers to symmetrization over indices with subtraction of the trace and anti-symmetrization, respectively. The first group of operators gives rise to the isovector axial, scalar and tensor charges $g_{A,S,T}^{u-d}$, while the second group are related to the isovector average quark momentum fraction $\avgx{-}{}$ and the helicity and transversity moments, $\avgx{-}{\Delta}$ and $\avgx{-}{\delta}$, respectively. \par

At zero-momentum transfer the form factor decomposition $W^X_{\mu_1...\mu_n}$ on the r.h.s. of Eq.~(\ref{eq:NMEs}) with appropriate choices of index combinations gives access to the desired observables from a ratio of spin-projected and gauge-averaged two- and three-point functions as given in Eqs.~(\ref{eq:2pt})~and~(\ref{eq:3pt})
\begin{equation}
 R^X_{\mu_1...\mu_n}(\tins, \tsep) = \frac{\langle C^X_{\mu_1...\mu_n}(\Gamma_3; \tins, \tsep) \rangle}{\langle C^\mathrm{2pt}(\Gamma_3; \tsep) \rangle} \,,
  \label{eq:ratio}
\end{equation}
in the limit of asymptotically large Euclidean time-separations $\tsep$ and $\tins$. The three-point functions are spin-projected with $\Gamma_3\equiv\Gamma_0\l(1+i\g{5}\g{3}\r)$ and the same is used for the two-point functions entering the ratio to optimally exploit correlations. \par

\subsection{Extraction of ground state NMEs}
In practice, source-sink separations of $\tsep\gg 1\fm$ are needed to adequately suppress excited state in the extraction of ground state matrix elements, thus a naive plateau analysis of the ratio for the accessible values of $\tsep \lesssim 1.5\fm$ is insufficient. Three-point functions at larger values of $\tsep$ cannot be computed with the desired statistical precision due to a severe signal-to-noise problem, which triggered the development of many, more sophisticated approaches to deal with excited states; for a review see Ref.~\cite{Ottnad:2020qbw}. \par

Here we make use of two versions of the summation method \cite{Maiani:1987by,Dong:1997xr,Capitani:2012gj} that is based on the sum of the ratio in Eq.~(\ref{eq:ratio}) over insertion times $S(\tsep)=\sum_{\tins=a}^{\tsep-a} R(\tins, \tsep)$. The first one is the original, ``plain'' summation method linear in $\tsep$
\begin{equation}
 S(\tsep, \tex=a) = \mathrm{const} + M_{00} \tsep \,,
\label{eq:summation_method_plain}
\end{equation}
that is employed as a cross-check for the second one, which is based on the two-state truncation of $S(\tsep, \tex=a)$ as described in Ref.~\cite{Ottnad:2021tlx}
\begin{equation}
 S(\tsep, \tex=a) = M_{00}(\tsep - a) + 2 \tilde{M}_{01} \frac{ e^{-\Delta a} - e^{-\Delta\tsep} }{1-e^{-\Delta a}} \,,
 \label{eq:summation_method_2state}
\end{equation}
and fitted simultaneously to all six isovector NMEs with the energy gap $\Delta$ between ground and first excited state as a common fit parameter to extract the ground state NMEs $M_{00}$. 

\begin{figure}[thb]
 \centering
 \includegraphics[totalheight=0.193\textheight]{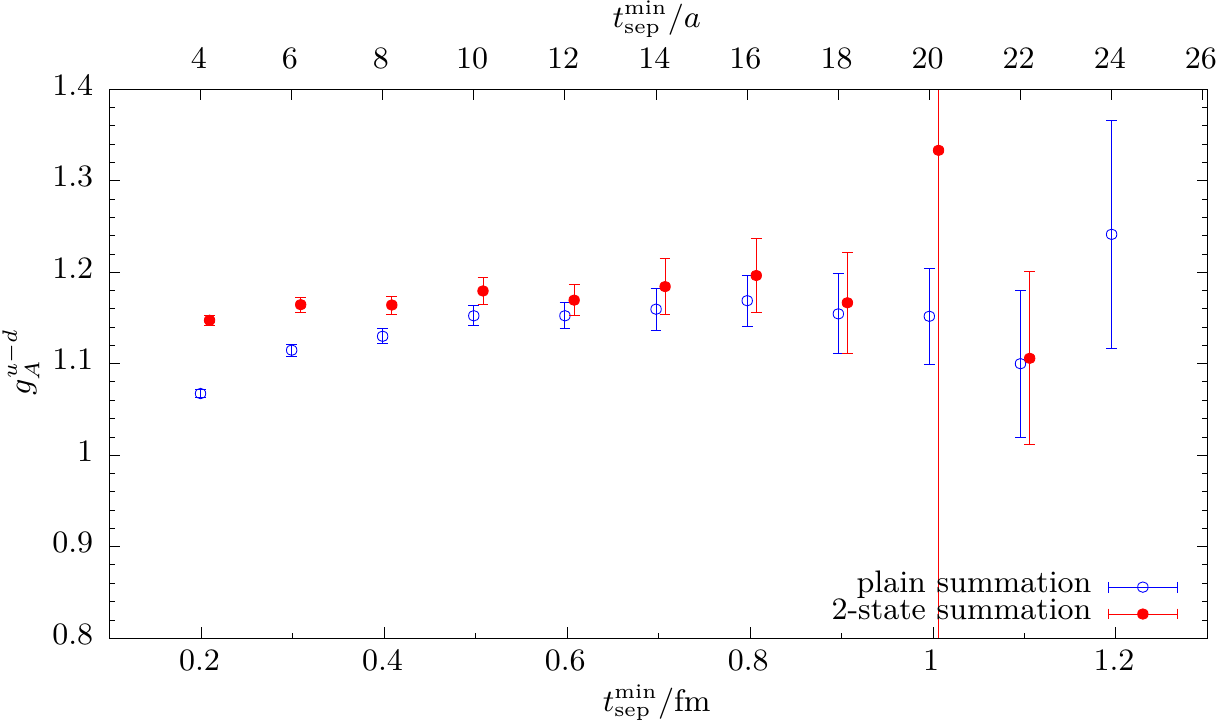}
 \includegraphics[totalheight=0.193\textheight]{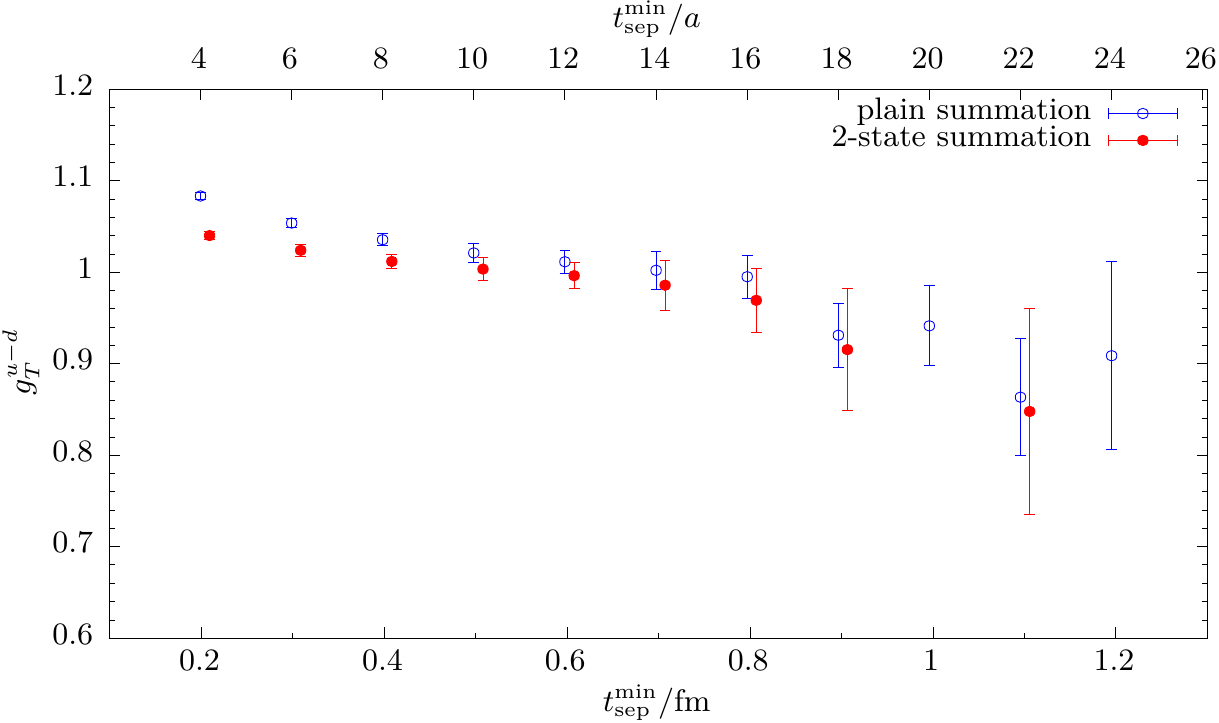} \\[3pt]
 \includegraphics[totalheight=0.193\textheight]{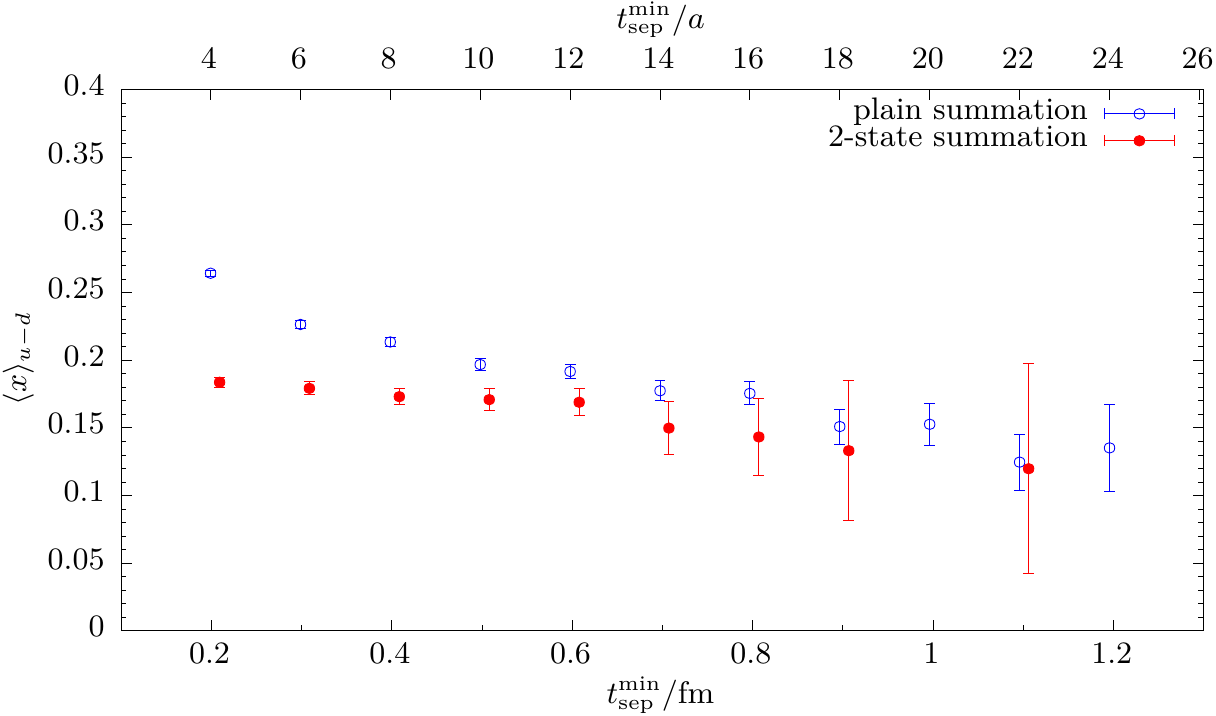}
 \includegraphics[totalheight=0.193\textheight]{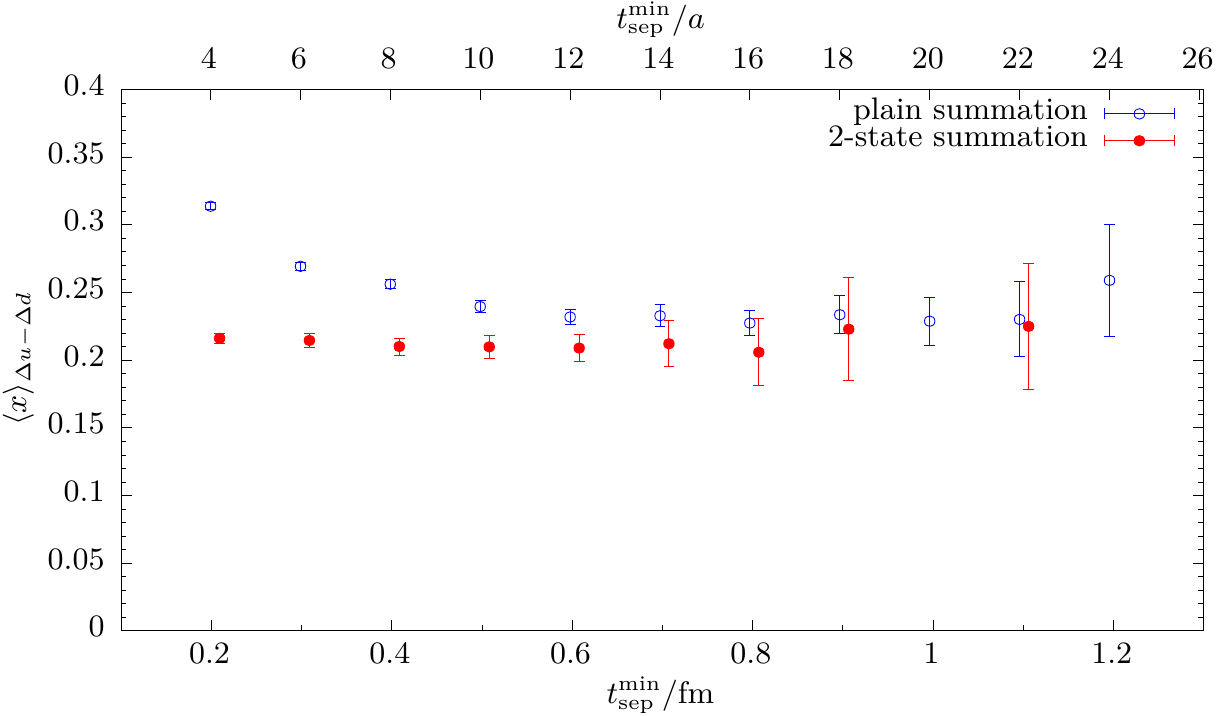}
 \caption{Results for $g_{A,T}^{u-d}$, $\avgx{-}{}$ and $\avgx{-}{\Delta}$ from plain and two-state summation method as a function of $\tsepmin$ on the E300 ensemble.}
 \label{fig:plain_vs_two_state_summation}
\end{figure}

The efficacy of both methods is readily assessed by comparing their respective results for the ground state NME as a function of $\tsepmin$. An example is shown in Fig.~\ref{fig:plain_vs_two_state_summation} for a selection of four observables on the E300 ensembles. We find that the two-state fits always exhibit a much more rapid convergence towards the ground state. The final choice of $\tsepmin$ is made such that $M_\pi\tsepmin\gtrsim0.7$ and $M_\pi\tsepmin\gtrsim0.5$ hold for the plain and two-state model, respectively. On some ensembles with large effective statistics $\tsepmin$ is increased further to obtain a reasonable fit quality. In general, we find that simultaneous two-state fits with a single gap lead to a good description of the data once the gap has converged. The gap typically approaches a value compatible with an $N\pi$ state, at least on ensembles with statistics sufficiently large to track it as a function of $\tsepmin$ before the signal is lost in noise. However, we observe a tension for low values of $\tsepmin$ between local and twist-2 NMEs. In particular, we find the matrix elements for $g_{A,T}^{u-d}$ to be clearly dominated by an $N\pi$ state even at the smallest values of $\tsepmin$ when fitted separately, whereas the leading contamination for the twist-2 NMEs seems to be generally dominated by a (much) heavier state. At larger values of $\tsepmin$ as used in our fits this tension becomes increasingly washed out. Still, it may be worthwhile to investigate an ansatz with two separate (effective) gaps for local and twist-2 NMEs in the future. \par

\subsection{Physical extrapolations}
\begin{figure}[thb]
 \centering
 \includegraphics[totalheight=0.25\textheight]{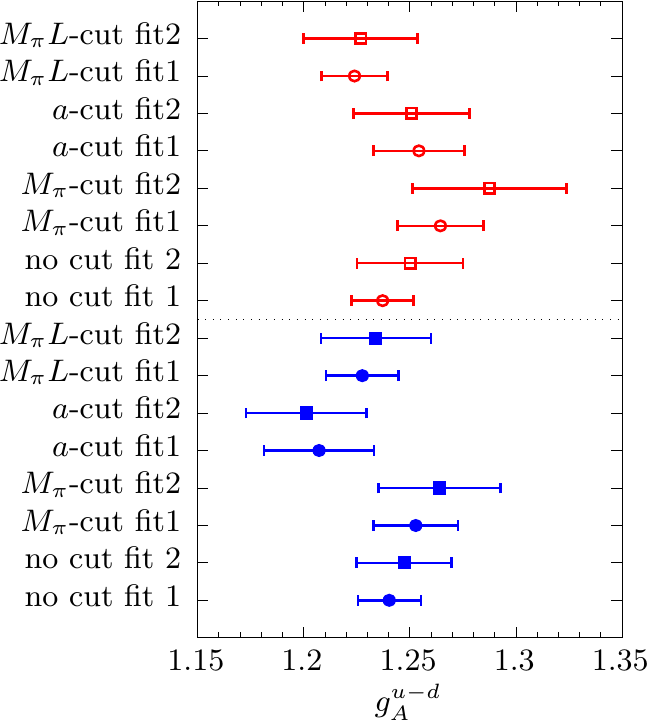}
 \includegraphics[totalheight=0.25\textheight]{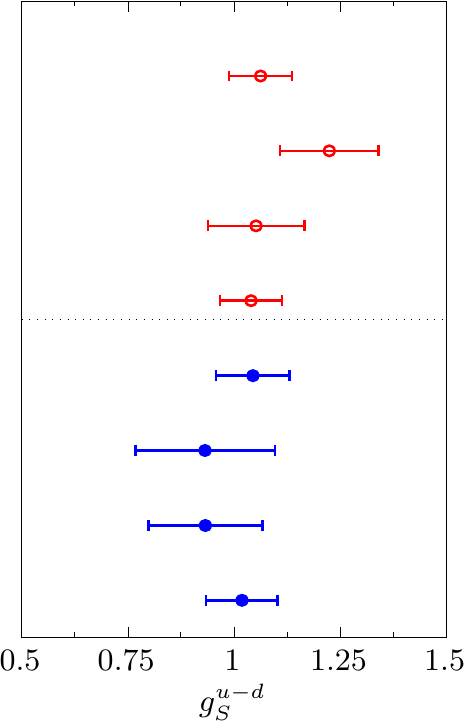}
 \includegraphics[totalheight=0.25\textheight]{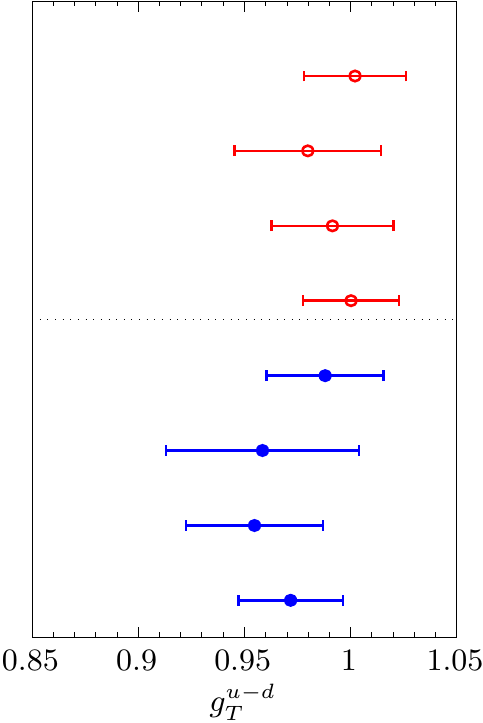} \\
 \includegraphics[totalheight=0.25\textheight]{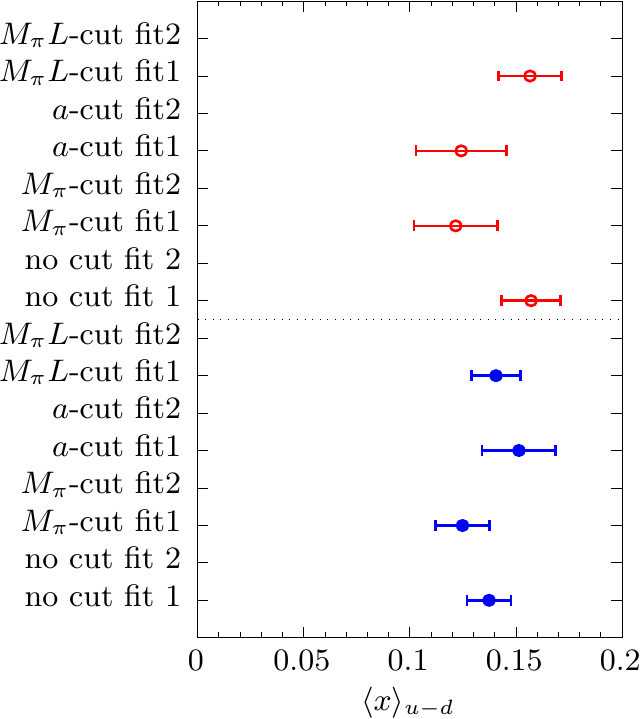}
 \includegraphics[totalheight=0.25\textheight]{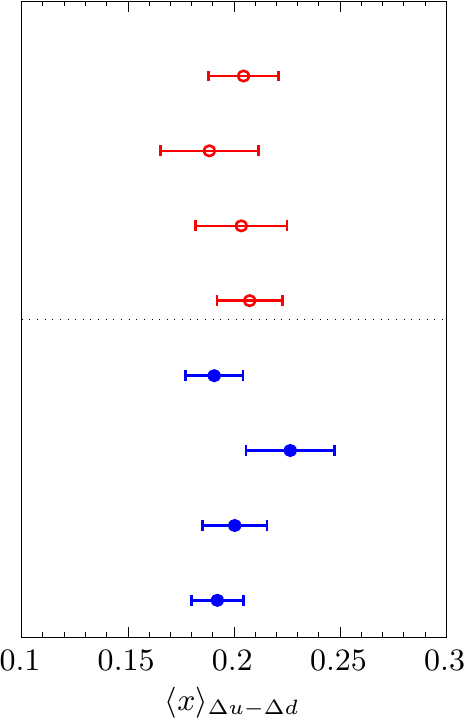}
 \includegraphics[totalheight=0.25\textheight]{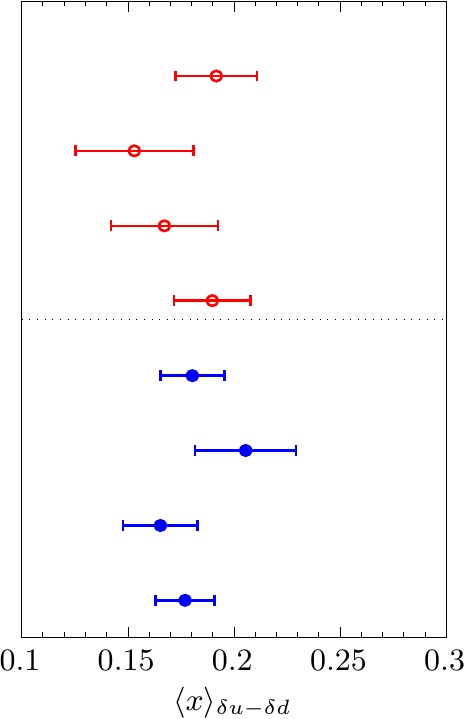}
 \caption{Overview of preliminary results for all six NMEs from different analysis procedures. Blue (red) symbols represent data obtained from the plain (two-state) summation method. Circles and boxes correspond to the two fit models for the physical extrapolation; see text. Three types of cuts ($M_\pi<300\mev$, $a<0.08\fm$, $M_\pi L>4$) have been applied to the data entering these fits, as indicated in the caption of the plot.}
 \label{fig:fit_models_overview}
\end{figure}

After extraction of the ground state NMEs $O(M_\pi, a, L)$ on individual ensembles, the physical extrapolation needs to be carried out. To this end we employ an ansatz inspired by NNLO $\chi$PT expression for the axial charge \cite{Kambor:1998pi}
\begin{equation}
 O(M_\pi,a,L) = A_O + B_O M_\pi^2 + C_O M_\pi^2 \log M_\pi + D_O M_\pi^3 + E_O a^{n(O)} + F_O \frac{M_\pi^2}{\sqrt{M_\pi L}} e^{-M_\pi L} \,,
 \label{eq:CCF_NME}
\end{equation}
where the leading lattice artifact is given by $n(O)=2$ for $O=g_{A,S}^{u-d}$ and $n(O)=1$ otherwise. The last term is included to correct for finite volume effects \cite{Beane:2004rf}, which can reach the few-percent level for $g_A^{u-d}$ depending on the box size; see also Refs.~\cite{Harris:2019bih,Ottnad:2021tlx}. The coefficient $C_O$ is known analytically, and in case of the axial charge it is given by
\begin{equation}
 C_{g_A}=\frac{-A_{g_A}}{(2\pi f_\pi)^2} \,,
 \label{eq:C_g_A}
\end{equation}
where $A_{g_A}=\chiral{g}_A^{u-d}$ is the isovector axial charge in the chiral limit and $f_\pi$ the pion decay constant. Therefore, $C_O$ is generally not a free parameter in the fit. However, fitting only the NLO expression without the $\mathcal{O}(M_\pi^3)$ term, one finds that the chiral logarithm imposes a curvature not observed in our data and fitting it with a free coefficient even yields the wrong sign. In case of the axial charge we thus consider two different fit models:
\begin{itemize}
 \item ``fit 1'': A NLO fit without the chiral logarithm, i.e. Eq.~(\ref{eq:CCF_NME}) with $C_O=D_O=0$.
 \item ``fit 2'': A (full) NNLO fit as given in Eq.~(\ref{eq:CCF_NME}) with $C_O$ expressed through Eq.~(\ref{eq:C_g_A}).
\end{itemize}
For all other observables we currently employ only the first model. To gain further insight into systematics we apply the same cuts to the data entering the physical extrapolations as previously used for $M_N$. An overview of preliminary results from all these variations is shown in Fig.~\ref{fig:fit_models_overview} for each observable. First of all, we observe an overall good agreement between results using data from the plain and two-state summation method, indicating that residual effects due to excited-state contamination are mostly under control. Secondly, the differences between results from fit 1 and fit 2 for the same set of input data of $g_A^{u-d}$ are small compared to the variation due to cuts. On the other hand, applying the $M_\pi<300\mev$ cut seems to favor larger results, hinting at some possible tension in the chiral extrapolation. Altogether the model spread is not negligible, which is particularly true for the statistically precise $g_A^{u-d}$, but much less so for e.g. $g_T^{u-d}$. However, the fit qualities indicated by the resulting $p$-values can vary strongly for different cuts and models on the same observable. Therefore, we intend to perform model averages for the final analysis to quantify systematic effects. \par

\section{Summary and outlook} \label{sec:outlook}
In this proceedings contribution we have presented the current status of our analysis of isovector NMEs at vanishing momentum transfer as well as the nucleon mass on the same set of data. This set of ensembles and measurements can now be considered complete for this study and the final analysis is underway. From the physical extrapolation of the nucleon mass data we find good agreement with the experimental value within the scale setting uncertainty, which corroborates the $t_0$ value from the dedicated scale setting study in Ref.~\cite{Bruno:2016plf}. However, for the final $M_N$ and NME analysis we intend to switch to a more recent and precise determination of the scale $t_0$, i.e. the FLAG average value in Ref.~\cite{FlavourLatticeAveragingGroupFLAG:2021npn}. \par 

Regarding the isovector NME analysis, we still plan to further refine our excited-state analysis, e.g. by allowing different gaps for matrix elements of local and twist-2 operators in the simultaneous fit for the two-state model for the summation method. It remains to be seen if this will have an impact on the physical extrapolation of the most precise NME(s) and if it further reduces the model spread observed in some of the cuts and fit variations for $g_A^{u-d}$ as shown in Fig~.\ref{fig:fit_models_overview}. The caveat here is that cuts in pion mass or the lattice spacing that are intended to test the robustness of the chiral and continuum extrapolation, may actually also reveal residual excited-state effects if these are e.g. enhanced by smaller pion masses or depend on the available $\tsepmin$ resolution. At any rate, for the final analysis we aim at obtaining physical results with fully controlled and quantified systematics from model averages. \par

\section*{Acknowledgments}
Two of the authors (GvH and KO) are supported by the Deutsche Forschungsgemeinschaft (DFG, German Research Foundation) through project HI~2048/1-2 (project No.~399400745). Calculations have been performed on the HPC clusters Clover at the Helmholtz-Institut Mainz and Mogon II and HIMster-2 at Johannes-Gutenberg Universit\"at Mainz. We gratefully acknowledge the support of the John von Neumann Institute for Computing and Gauss Centre for Supercomputing e.V. (http:www.gauss-centre.eu) for projects CHMZ21, CHMZ36 and NUCSTRUCLFL. We thank our colleagues in the CLS initiative for sharing ensembles.

\end{document}

%% file: newcommands.tex
\renewcommand{\l}{\left}
\renewcommand{\r}{\right}
\newcommand{\g}[1]{\gamma_{#1}} 
\newcommand{\DB}[1]{\stackrel{\leftarrow}{D}_{#1}} 
\newcommand{\DF}[1]{\stackrel{\rightarrow}{D}_{#1}} 
\newcommand{\DBF}[1]{\!\stackrel{\leftrightarrow}{D}_{#1}} 

\newcommand{\bra}[1]{\left< #1 \right|} 
\newcommand{\ket}[1]{\left| #1 \right>} 

\newcommand{\chiral}[1]{\mathring{#1}} 

\newcommand{\mev}{\,\mathrm{MeV}}
\newcommand{\fm}{\,\mathrm{fm}}

\newcommand{\phys}{\mathrm{phys}}
\newcommand{\stat}[1]{(#1)_\mathrm{stat}}
\newcommand{\sys}[1]{(#1)_\mathrm{sys}}
\newcommand{\tex}{t_\mathrm{ex}}
\newcommand{\tins}{t_\mathrm{ins}}

\newcommand{\tsep}{t_\mathrm{sep}}
\newcommand{\tsepmin}{t_\mathrm{sep}^\mathrm{min}}

\newcommand{\tseplo}{t_\mathrm{sep}^\mathrm{lo}}
\newcommand{\tsephi}{t_\mathrm{sep}^\mathrm{hi}}

\newcommand{\avgx}[2]{\langle x \rangle_{#2 u #1 #2 d}}